\documentclass[twocolumn,secnumarabic,amssymb, nobibnotes, aps, prl,nofootinbib,superscriptaddress]{revtex4-2}

\usepackage{orcidlink}

\usepackage[T1]{fontenc} 
\usepackage[utf8]{inputenc} 
\usepackage{microtype} 
\usepackage{lmodern} 
\usepackage{booktabs} 
\usepackage{mdframed} 
\usepackage{graphicx}

\usepackage{hyperref} \usepackage{color}
\definecolor{dark-red}{rgb}{0.4,0.15,0.15}
\definecolor{dark-blue}{rgb}{0.15,0.15,0.4}
\definecolor{medium-blue}{rgb}{0,0,0.5} \hypersetup{colorlinks,
  linkcolor={dark-red}, citecolor={dark-blue}, urlcolor={medium-blue}}

\usepackage{amsmath,amsfonts,amssymb}
\usepackage{eucal} 
\usepackage{mleftright} \mleftright
\usepackage{tensor} 
\usepackage{braket} 
\usepackage{bm} 
\usepackage{mathrsfs} 

\providecommand*{\pd}{\partial}
\renewcommand*{\pd}{\partial}

\begin{document}

\title{Superfluid memory effect}

\author{Kristan Jensen\,\orcidlink{0000-0002-5864-6416}}%
\email{kristanj@uvic.ca}
\affiliation{Department of Physics and Astronomy, University of Victoria, Victoria, BC V8W 3P6, Canada}
\affiliation{Erwin Schr\"odinger Int.\ Institute for Mathematics and Physics, University of Vienna}

\author{Alfredo Pérez\,\orcidlink{0000-0003-0989-9959}}%
\email{alfredo.perez@uss.cl}
\affiliation{Erwin Schr\"odinger Int.\ Institute for Mathematics and Physics, University of Vienna}
\affiliation{Centro de Estudios Científicos (CECs), Avenida Arturo Prat 514, Valdivia, Chile}
\affiliation{Facultad de Ingeniería, Arquitectura y Diseño, Universidad San Sebastián, sede Valdivia, General Lagos 1163, Valdivia 5110693, Chile}

\author{Stefan Prohazka\,\orcidlink{0000-0002-3925-3983}}%
\email{stefan.prohazka@univie.ac.at}
\affiliation{Erwin Schr\"odinger Int.\ Institute for Mathematics and Physics, University of Vienna}
\affiliation{University of Vienna, Faculty of Physics, Mathematical Physics, Boltzmanngasse 5, 1090, Vienna, Austria}


\begin{abstract}
We identify a memory effect produced by phonon scattering in a superfluid: a localized scattering event generates a far-field pressure pulse with vanishing time integral but a nonzero first temporal moment, corresponding to a permanent shift in a prepotential for the far-field velocity. The effect is controlled by the soft factor for phonon emission, including corrections from nonlinear dispersion. In superfluid $^4$He, scattering macroscopic phonon wave packets produces a signal whose estimated magnitude may lie within experimental reach, thereby providing a laboratory analogue of electromagnetic and gravitational memory.
\end{abstract}

\maketitle

\emph{Introduction.}
\label{sec:introduction}
In this Letter we consider phonon scattering in a superfluid. Our main result is that a localized scattering event leaves a permanent, potentially observable imprint on a variable constructed from the velocity field far from the scattering. We call this effect ``superfluid memory,'' in analogy with electromagnetic~\cite{Bieri:2013hqa}
and gravitational memory~\cite{Zeldovich:1974gvh,Braginsky:1985vlg,braginsky1987gravitational,Christodoulou:1991cr,Thorne:1992sdb}. It follows from the Goldstone effective field theory (EFT) of the
superfluid phonon.

The leading self-interaction produces a finite soft factor for phonon emission. Its form depends on low-energy couplings, including those controlling nonlinear phonon dispersion. We compute these corrections and show that the resulting soft factor determines the memory encoded in the far-field velocity profile. Nonlinear dispersion regulates the collinear enhancement of soft phonon emission.

The superfluid memory is determined by the momenta of the incoming and outgoing phonons together with the low-energy couplings of the fluid. It provides a tabletop analogue of electromagnetic and gravitational memory, for which the corresponding infrared processes are soft photon and soft graviton emission. Neither electromagnetic nor gravitational memory has yet been observed, although networks of second-generation ground-based detectors and planned space-based interferometers may detect the gravitational displacement memory produced by binary black hole mergers within the next decade~\cite{Grant:2022bla,Inchauspe:2024ibs}.

Superfluid memory, by contrast, may already be accessible in the laboratory. Unlike gravitational memory, however, its magnitude depends on several material-dependent low-energy couplings rather than on a single fundamental interaction. To assess its experimental viability, we estimate the effect in pressurized superfluid $^4$He at low temperature and consider the scattering of macroscopic phonon wave packets, whose large occupation numbers can amplify the signal. Our estimates suggest that the resulting memory may lie within experimental reach.
  
\emph{Superfluid EFT.}
\label{sec:he-ii-effective}
The zero-temperature, long-wavelength dynamics of a non-relativistic superfluid is controlled by the Goldstone field $\pi$ associated with the spontaneous breaking of a global $U(1)$ symmetry. The corresponding EFT is invariant under the shift symmetry $\pi\to\pi+c$ as well as translations and spatial rotations.  The shift symmetry implies that the Goldstone is derivatively coupled, and the Lagrangian can be expanded in powers of the Goldstone and of derivatives with the result (throughout we work in $\hbar=1$ units)
\begin{align}
\nonumber
	\mathcal{L} &= \frac{1}{2c_{s}^{2}}\dot{\pi}^{2}
- \frac{1}{2}(\partial_{i}\pi)^{2} + \gamma_1(\pd_{i}^{2}\pi)^{2} + \frac{\gamma_2}{c_s^2} (\partial_i\dot{\pi})^2 + \frac{\gamma_3}{c_s^4} \ddot{\pi}^2
	\\
	&\quad 
	+ g_1\dot{\pi}(\partial_{i}\pi)^{2} + \frac{g_2}{c_s^2}\dot{\pi}^{3} +\hdots\,.
\label{eq:L_general}
\end{align}
The dots indicate terms with more derivatives and/or more powers of $\pi$. Ignoring the interactions, the phonon propagates with a dispersion relation
\begin{equation}
	E = \pm c_s p\left( 1 - \gamma p^2 + O(p^4)\right)\,, \qquad \gamma = \gamma_1 + \gamma_2 + \gamma_3\,.
    \label{E:dispersion}
\end{equation}
The couplings $\gamma_i$ and $g_i$ are not all independent. Scattering amplitudes and dispersion depend on $\gamma$ and $g = g_1 + g_2$. The sign of $\gamma$ controls whether low-energy phonons can undergo Beliaev decay~\cite{Landau:1949xrz,Khalatnikov:Superfl,beliaev1958energy}. When $\gamma>0$, a low-energy $1\to 2$ process is kinematically forbidden, while for $\gamma<0$ it is allowed. We consider $\gamma>0$.

When the superfluid is Galilean invariant, as in $^4$He, Galilean symmetry imposes constraints on the couplings and furthermore relates some of them to the zero-temperature equation of state $P(\mu)$. To understand these relations, suppose the underlying particles have a mass $m$ and let $\theta\sim \theta+2\pi$ be the Goldstone mode. Galilean transformations by a boost $v^i$ with $x^i \to x^i - v^i t$ act as
\begin{equation}
	\delta_v\theta = t\,v^i\partial_i \theta - m\,v^{i}x_{i}\,,
\label{eq:Galilei_boost}
\end{equation}
so that 
\begin{align}
	X &= \mu -\dot{\theta} - \frac{1}{2m}(\partial_{i}\theta)^{2}\,, \quad \partial_i \partial_j \theta\,, 
\\
	D_t X & = \left( \partial_t + \frac{\partial^i\theta}{m}\partial_i\right)X\,, \qquad \partial_i X\,,
\end{align}
with $\mu$ the chemical potential, are boost-invariant and so can be used to construct a Galilean-invariant Lagrangian.

At leading order in derivatives, the EFT is fixed by the equation of state $P(\mu)$ to be~\cite{Greiter:1989qb,Son:2002zn,Son:2005rv}
\begin{equation}
	\mathcal{L} = P(X) + (\text{higher derivative terms})\,. 
\end{equation}
To relate this description to~\eqref{eq:L_general} we rescale $\theta$ as
\begin{equation}
	\pi = \sqrt{\frac{P'(\mu)}{m}}\,\theta\,,
\end{equation}
	and expand around $X=\mu$. This gives~\eqref{eq:L_general} with
\begin{equation}
	c_{s}^{2} = \frac{P'}{mP''}\,, 
	\quad 
	g_1 = \frac{\sqrt{m}}{2}\frac{P''}{(P')^{3/2}}\,, 
	\quad 
	g_2 = -\frac{c_s^2}{3!} \frac{m^{3/2}P'''}{(P')^{3/2}}\,,
\label{eq:couplings}
\end{equation}
fixed by $P(\mu)$ and its derivatives, and $\gamma$ is an independent higher-derivative coupling.

\emph{Phonon scattering and soft factor.}
\label{sec:phon-scatt-soft}
The Feynman rules following from~\eqref{eq:L_general} give the phonon
propagator
\begin{equation}
	D(E,\bm{p}) =
	\frac{i}{E^{2}/c_{s}^{2}-\bm{p}^{2}+2\gamma\,\bm{p}^{4}+i\varepsilon}\,,
\label{eq:propagator}
\end{equation}
and the cubic vertex (with $i=1,2,3$ labeling the legs)
\begin{equation}
	V_{3} =- 2g_1\!\!\sum_{\text{cycl}(i,j,k)}\! E_{i}\,(\bm{p}_{j}\!\cdot\!\bm{p}_{k}) -\frac{6g_2}{c_s^2}\,E_{1}E_{2}E_{3}
\,.
\label{eq:vertices}
\end{equation}

Consider a scattering process involving $N$ hard phonons with momenta	$\bm{p}_{\alpha}=p_{\alpha}\bm{n}'_{\alpha}$. We then study the same process accompanied by the emission of a soft phonon with energy $\omega\to 0$ and momentum $\bm{q}=\frac{\omega}{c_{s}}\bm{n}$ where $\bm{n}$ is a unit vector.  We retain the leading cubic interaction $g$ and the leading nonlinear correction to the dispersion, which becomes important in the collinear limit.

In this regime the tree-level $(N+1)$-point amplitude factorizes as
\begin{equation}
\label{eq:scatter}
	\lim_{\omega \to 0} \mathcal{A}_{N+1} \left(\{E_{\alpha},\bm{p}_{\alpha}\};(\omega,\bm{n})\right)
= \mathcal{S}\, \mathcal{A}_{N} \left(\{E_{\alpha},\bm{p}_{\alpha}\}\right) +O(\omega)\,,
\end{equation}
where the soft factor $\mathcal{S}$ is obtained by attaching the soft	phonon to each external hard leg. We find
\begin{equation}
\label{eq:soft_factor}
	\mathcal{S} =
	\sum_{\alpha=1}^{N}
	\frac{-3i c_{s} \eta_{\alpha} p_{\alpha} g}
	{1-\bm{n}\cdot \bm{n}'_{\alpha}-\gamma p_{\alpha}^{2}  \left(1-4 \bm{n}\cdot \bm{n}'_{\alpha}\right)}\,,
\end{equation}
with $\eta_{\alpha}=+1$ ($-1$) for outgoing (incoming) hard phonons. The analogous factor with $\gamma=0$ was found in~\cite{Green:2022slj,Cheung:2023qwn}. We keep the dispersive correction in the nearly on-shell intermediate propagator, while evaluating the remaining factors at leading order in the derivative expansion.

There are two features of~\eqref{eq:soft_factor} worth noting. First, the leading behavior of $\mathcal{S}$ is $O(\omega^0)$, so that the soft limit is finite and non-vanishing, in contrast with the Adler zero~\cite{Adler:1964um} of relativistic Goldstone theories. Here the soft factor is generated by the cubic interaction $g$ allowed in a non-relativistic superfluid. Second, the factor depends on higher-derivative EFT data and will in general receive further gradient corrections. We therefore refer to it as a soft \emph{factor} rather than a soft theorem. Whether it also receives loop corrections is an open question.

In particular, the correction to dispersion plays a physically important role. For $\gamma=0$ the denominator reduces to $1-\bm n\cdot \bm n'_{\alpha}$, exhibiting a collinear divergence as $\bm n\to \bm n'_{\alpha}$, which is regulated by $\gamma>0$ in the regime $1-\bm n\cdot \bm n'_{\alpha} \sim \gamma p_{\alpha}^2$. Higher-derivative interactions will give further corrections which we do not consider.

\emph{Memory effect.}
\label{sec:memory-effect}
Now we show how the soft factor in phonon scattering implies the permanent displacement of a quantity constructed from the far-field velocity, which we call the superfluid memory effect. Our approach is similar to that of~\cite{Strominger:2014pwa}.

At large distances from the scattering region and to leading order in the phonon interaction, soft phonons are described by the standard mode decomposition of free massless fields.  Introducing retarded coordinates $u = t - r/c_s$ and performing a saddle-point approximation at large $r$~\cite{Strominger:2014pwa}, one finds
\begin{equation}
	\pi = \frac{\bar{\pi}(u,\bm{n})}{r} + \cdots \,,
\label{eq:pi_asymptotic}
\end{equation}
with (for $\omega_q = c_s q$)
\begin{equation}
	\bar{\pi}(u,\bm{n})
	= \frac{1}{8i\pi^{2}}\int_{0}^{\infty} d\omega_{q}
	\Big(
	a_{\omega_{q} \bm{n}} e^{-i\omega_{q} u}
	- a^{\dagger}_{\omega_{q} \bm{n}} e^{i\omega_{q} u}
\Big).
\label{eq:bar_pi}
\end{equation}
Interactions modify the far-field asymptotics, through logarithmic terms analogous to those in~\cite{Fuentealba:2024lll,Briceno:2025cdu}, at higher orders than we consider here.

The velocity field is related to the Goldstone $\pi$ as $\bm v = 2g_1 c_s^2 \bm \nabla \pi$. At large $r$ it is radial with $\bm v = \bar{v}^r \bm n/r$ and
\begin{equation}
\label{E:constitutive}
	\bar{v}^r = -2g_1 c_s \partial_u\bar{\pi}\,.
\end{equation}
Because the soft factor $\mathcal{S}$ is regular as $\omega\to 0$,
\begin{equation}
\label{eq:zero_average}
	\boxed{\int_{-\infty}^{\infty} du \,\bar{v}^r = 0\,,}
\end{equation}
or equivalently $\bar{\pi}(u=\infty) - \bar{\pi}(u=-\infty) = 0$. This implies that we can write
\begin{equation}
	\bar{v}^r = \frac{\partial^2\mathcal{M}}{\partial u^2}\,,
\end{equation}
where we have defined $\mathcal{M}$ to be the velocity \emph{prepotential} which implicitly obeys $\partial_u \mathcal{M}\to 0$ at early and late times.

We now show that the soft factor encodes a permanent displacement in $\mathcal{M}$ under scattering. Although $\mathcal M$ is defined only up to an additive constant, this displacement is unambiguous. To do so, we use standard distributional identities to show
\begin{equation}
	\int_{-\infty}^{\infty} du \,u \partial_u \bar{\pi}(u,\bm n) = \lim_{\omega\to 0^+}\partial_{\omega} \left( \frac{i}{8\pi}\omega\left( a_{\omega \bm n} - a^{\dagger}_{\omega \bm n}\right)\right)\,.
\end{equation}
Inserting this expression between scattering states and using the soft factorization~\eqref{eq:scatter} $\langle \text{out} |a_{\omega \bm n}S|\text{in}\rangle = \mathcal{S}\langle \text{out}|S|\text{in}\rangle + O(\omega)$, we can write
\begin{align}
	\lim_{\omega \to 0^{+}}\pd_{\omega}[\omega \braket{\mathrm{out}|a(\omega \bm{n})S|\mathrm{in}} ] = \mathcal{S} \braket{\mathrm{out}|S|\mathrm{in}}\,,
\end{align}
and so
\begin{equation}
	\int_{-\infty}^{+\infty}du \, u \partial_{u}\bar{\pi}(u,\bm{n}) = \frac{i}{8\pi}\mathcal{S}\,.
\label{eq:Delta_soft}
\end{equation}
In this expression $\bar{\pi}$ is the far field expectation value of the corresponding operator conditioned on the specified hard in- and out-states. It is real since $\mathcal{S}$ is imaginary.

Relating $\bar{\pi}$ to the outgoing velocity and using
\begin{equation}
	\Delta \mathcal{M}=\mathcal{M}(u=\infty) - \mathcal{M}(-\infty)=-\int_{-\infty}^{\infty} du\,u  \bar{v}^r\,,
\end{equation}
we then obtain our central result, the memory effect
\begin{equation}
\label{eq:main_result}
	\boxed{ 
	\Delta \mathcal{M} = 
	\frac{1}{4\pi}\!\sum_{\alpha=1}^{N}\!
	\frac{3 c_{s}^{2}g g_1 \eta_{\alpha}p_{\alpha}}
	{1-\bm{n}\!\cdot\! \bm{n}'_{\alpha}-\frac{\gamma p_{\alpha}^{2}}{\hbar^2}(1\!-\! 4 \bm{n}\!\cdot\! \bm{n}'_{\alpha})}\,.
	}
\end{equation}
where we have restored $\hbar$. The displacement is fully determined at this order by the EFT couplings and the hard phonon momenta, with $\bm{n}\cdot\bm{n}'_\alpha = \cos\vartheta_{\alpha}$ the angle between the observation and hard phonon directions.

These results also constrain the far-field soft pressure pulse generated by scattering, which we expect to be easier to measure. Using linear superfluid hydrodynamics we relate velocity perturbations to pressure fluctuations,
\begin{equation}
	\dot{\bm v} = - \frac{1}{m n}\bm \nabla \delta P\,,
\end{equation}
where $n$ is the number density. In terms of $\delta P = \frac{\delta\bar{P}}{r} + \hdots$ we have
\begin{equation}
	\delta \bar{P} = c_s m n \bar{v}^r\,,
\end{equation}
so that~\eqref{eq:zero_average} and~\eqref{eq:main_result} become
\begin{equation}
\label{eq:pressure_moments}
	\int_{-\infty}^{\infty} du\,\delta\bar{P} = 0\,, \qquad \int_{-\infty}^{\infty}du\,u \delta \bar{P} = - c_s m n \Delta \mathcal{M}\,.
\end{equation}

\begin{figure}
  \includegraphics[width=\linewidth ,keepaspectratio]{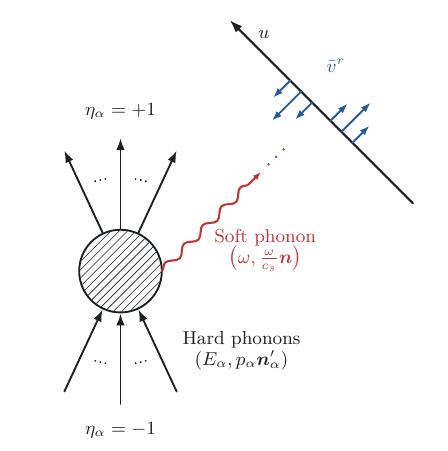}
\caption{Summary of the central results~\eqref{eq:zero_average} and~\eqref{eq:main_result}. Hard phonon scattering produces a transient radial velocity pulse far from the scattering region. Because the velocity pulse averages to zero,~\eqref{eq:zero_average}, it may be written as $\bar{v}^r = \frac{\partial^2\mathcal{M}}{\partial u^2}$, and the scattering implies a permanent displacement in $\mathcal{M}$, the superfluid memory~\eqref{eq:main_result}. There are similar results~\eqref{eq:pressure_moments} for the pressure far from the scattering.\label{fig:overview}} 
\end{figure}

Both predictions~\eqref{eq:zero_average} and~\eqref{eq:main_result} are in principle testable. The structure mirrors that of electrodynamics and general relativity, where soft theorems at successive orders in $\omega$ produce a hierarchy of memory effects (cf., e.g.,~\cite{Hamada:2018cjj}). In the superfluid case, the soft factor starts at $O(\omega^0)$ rather than $O(\omega^{-1})$, so the zeroth moment vanishes~\eqref{eq:zero_average} and the leading nontrivial memory is the first moment~\eqref{eq:main_result}.

\emph{Numerical estimates.}
\label{sec:numer-estim-meas}
We estimate the size of this effect in pressurized $^4$He and assess whether it may be experimentally detectable. The regime $\gamma>0$ is realized when the pressure and number density are $P\gtrsim14\,\mathrm{bar}$, $n\gtrsim0.0245\, \text{\AA}^{-3}$. We use the empirical equation of state~\cite{McCarty:Superfluid}
\begin{equation}
	P = b\Big[a_{1}(n-\bar{n})
	+ a_{2}(n-\bar{n})^{2}
	+ a_{3}(n-\bar{n})^{3}\Big] \, .
\label{eq:EOS}
\end{equation}
When $P$ is measured in bars and the number density $n$ in $\mathrm{mol}/\mathrm{liter}$ the fit parameters are
\begin{align*}
	b&=1.01325\,, &  a_{1}&=2.281877372\,, &  a_{2}&=0.16820886\,, 
	\\
	\bar{n}&=36.27877\,, & a_{3}&=0.005277884968 \,.
\end{align*}
We consider $n\approx 0.0255\,\text{\AA}^{-3}$ at which $P\approx 21.5$ bar, $\gamma\approx 0.311\,\text{\AA}^{2}$~\cite{Beauvois_2019} and $c_s\approx 352$ m/s. The coupling constants can then be determined from Eq.~\eqref{eq:couplings}, where the mass of the ${}^4$He isotope in atomic mass units is $m=4.0026$. 

We would like to consider scattering processes which are within the regime of perturbation theory in superfluid EFT, in particular where scattering can be approximated as $2\to 2$ and infrequent, nonlinear dispersion is a small effect, and temperature is sufficiently small as to ignore damping. At this number density this can be achieved at $k \approx 0.1 \,\text{\AA}^{-1}$, with a single-phonon energy $E_k \approx 3.7\times 10^{-23}$ J. Using the Landau-Khalatnikov $2\leftrightarrow 2$ damping rate we estimate an attenuation length $\sim 87$ m at a temperature $T=0.1$ K at this density and wave number.

The scattering of a few hard phonons at this energy is unlikely and the resulting pressure fluctuations are impracticably small to measure. For this reason, we consider the scattering of macroscopic phonon beams. The number of scattering events is enhanced by the occupation numbers of the incident beams and crucially the memory~\eqref{eq:main_result} is an additive effect for which the different scattering events contribute constructively. 

We consider back-to-back incident pulses peaked at wavenumber $k=0.1 \,\text{\AA}^{-1}$, with cross-sectional area of $A_{\perp} = 1$ mm$^2$ over a duration $\tau = 10^{-7}\,$s, corresponding to an energy of $10^{-8}$ J and occupation number $N = 2.7\times 10^{14}$ per pulse. 
This is qualitatively similar to the ``phonon sheets'' created in~\cite{10.1063/10.0043138}.

The resulting pressure memory is
\begin{equation}
\label{E:averagedPressureMemory}
	-\int_{-\infty}^{\infty} du \,u \delta\bar{P}(u,\bm n) = \frac{N^2}{A_{\perp}} \int d\Omega \frac{d\sigma}{d\Omega} c_s m n \Delta \mathcal{M}(\bm n;\Omega)\,,
\end{equation}
where $\Delta\mathcal{M}$ is the memory observed along axis $\bm n$ for a $2\to 2$ process with the two incoming hard phonons drawn from the incident beams and the outgoing phonons emitted back-to-back at angle $\Omega$. In this regime we can borrow the methods of Landau and Khalatnikov, summarized in~\cite{Kurkjian_2017} for the differential cross section,
\begin{align}
\begin{split}
	\frac{d\sigma}{d\Omega} &= \frac{m^2 c_s^4k^2}{2\pi^2 \hbar^2 n^2 v^2} |\mathcal{A}|^2\,,
	\\
	\mathcal{A}& = \frac{1}{16} \left( \frac{E_k}{mc_s^2}\right)^2 \left(\Sigma_F -18 \Lambda_F^2 - 2 \cos^2(\theta)\right)\,.
\end{split}
\end{align}
Here $v\approx 349$ m/s is the group velocity and 
\begin{equation}
    \Lambda_F = \frac{n}{3}\frac{\partial^2\mu}{\partial n^2}\left( \frac{\partial\mu}{\partial n}\right)^{-1} = \frac{g_2}{g_1}\,, \quad \Sigma_F = \frac{n^3}{mc_s^2}\frac{\partial^3\mu}{\partial n^3}\,.
\end{equation}
The amplitude follows from their Eq.~(111) upon using the kinematics of elastic $2\to 2$ scattering. This result  neglects nonlinear dispersion, which represents a small correction. At this density the equation of state gives $\Sigma_F -18 \Lambda_F^2 \approx - 20.9$. This gives a total cross section $\sigma \approx 3.64\times 10^{-4} \,\text{\AA}^2$ and so a scattering probability of $\sigma \tfrac{N}{A_{\perp}} \approx 10^{-3}$ per incident phonon, so our assumption of independent $2\to 2$ events is self-consistent.

\begin{figure}
\includegraphics[width=\linewidth ,keepaspectratio]{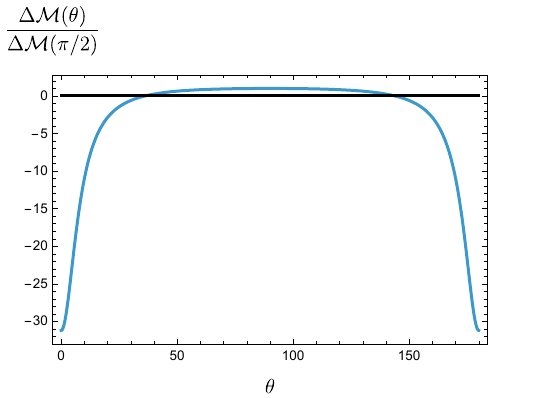}
\caption{The predicted angular profile of the memory discussed in the main text as a function of angle $\theta$ away from the beam axis, relative to the memory at $\theta = \pi/2$. The pressure memory follows the same curve. \label{F:memory-plot}}
\end{figure}

Suppose that the outgoing pressure pulse is resolved by a detector a distance $r$ away from the scattering and over a timescale $\tau_{\rm eff}$, then after averaging over the final state distribution as in~\eqref{E:averagedPressureMemory} we find that the resulting pressure pulse has the characteristic size
\begin{equation}
	\delta P_{\rm obs} \sim (-0.1, + 0.003)\,\text{Pa}\left( \frac{1 \text{ cm}}{r}\right) \left( \frac{\tau}{10^{-7} \,\text{s}}\right)^2 \left( \frac{10^{-6} \,\text{s}}{\tau_{\rm eff}}\right)^2\,,
\end{equation} 
where the quoted values are at collinearity and perpendicular to the beam axis, respectively. So for a pulse of length $\tau = 10^{-7}$ s, a detector placed $r=5$ cm away with a resolution $\tau_{\rm eff} = 3\,\mu$s the resulting signal is macroscopically appreciable, $\delta P_{\rm obs} \sim (-0.0024, + 0.000075)$ Pa. We plot the full angular profile in
Fig.~\ref{F:memory-plot}.

\emph{Discussion.}
\label{sec:discussion}
Starting from the EFT description of a non-relativistic superfluid, we have derived a memory effect in low-energy phonon scattering. To obtain it we first derived the soft factor~\eqref{eq:soft_factor} describing the emission of a soft phonon in tree-level processes. When the soft phonon has a frequency $\omega$, the factor is finite as $\omega\to 0$. It depends on the leading cubic interaction $g$ and the leading correction to linear dispersion $\gamma$, generalizing the results of~\cite{Green:2022slj,Cheung:2023qwn}. This soft factor constrains the temporal moments of the velocity field far from the scattering. The radial velocity averages to zero~\eqref{eq:zero_average}, and the first moment encodes a memory effect, the permanent displacement in a velocity prepotential~\eqref{eq:main_result}. This memory depends on the angle relative to hard in- and outgoing phonons, is peaked near collinearity with them, and the correction $\gamma$ to dispersion regulates the collinear limit.

We determined the various couplings from data for pressurized superfluid $^4$He, for which we estimated the strength of the memory effect. We proposed to measure it in the scattering of macroscopic phonon beams where the effect is enhanced by large occupation numbers. The velocity moments can be traded for those of pressure fluctuations, which seem easier to detect, and for which we predict distinctive and macroscopic results.

A memory effect due to sound waves has been discussed previously in~\cite{deAguiarAlves:2025vfu}. Our results are rather different. In particular we have accounted for the derivative couplings appropriate to a superfluid, essential both for our main results~\eqref{eq:zero_average} and~\eqref{eq:main_result} and our numerical estimates, through the differential cross section appropriate for phonons with momentum-suppressed interactions.

More broadly, our results open a new avenue for studying memory effects in condensed-matter and many-body systems as well as opening interesting theoretical questions. Usually soft theorems and memory effects are equivalent to asymptotic symmetries of the $S$-matrix. Is there such an equivalent symmetry in phonon scattering? Moreover are there analogous memory effects when scattering rotons, vortices, or impurities? What are the effects of quantum field theory loops on the soft physics? And finally, are there observable memory effects to be discovered in other arenas with a Goldstone description like those considered in~\cite{Nicolis:2015sra,Cheung:2023qwn}?

\emph{Acknowledgments.}
We thank M.~Briceno, R.~Emparan, H.~A.~Gonzalez and T. Maedler for valuable discussions. The authors used OpenAI ChatGPT (GPT-5.6 Sol) to develop a preliminary version of the section ``Numerical estimates.’’ We thank the participants of the Carrollian Physics and Geometry Workshop in Brussels where this work was presented and for useful discussions. We acknowledge support from the Erwin Schrödinger International Institute for Mathematics and Physics (ESI) where some of the research was undertaken during the ESI Research in Teams Programme. KJ was supported in part by an NSERC Discovery Grant. The research of AP is supported by the ANID Fondecyt grant 1260427. SP acknowledges support from the ChatGPT for Academic Researchers programme.

\bibliography{bibl}

\end{document}